\documentclass[prl,twocolumn]{revtex4}%
\usepackage{amsmath}%
\usepackage{amsfonts}%
\usepackage{amssymb}%
\usepackage{graphicx}
\begin{document}

\title{Photon wave function and position eigenvectors}
\author{Margaret Hawton}
\affiliation{Department of Physics, Lakehead University, Thunder Bay, ON, Canada, P7B 5E1}

\begin{abstract}
One and two photon wave functions are obtained by projection onto a basis of
simultaneous eigenvectors of the position and number operators.

\end{abstract}
\maketitle

The current interest in entanglement and its application to quantum
communications has rekindled the debate on the nature of the photon wave
function \cite{Raymer,LapaireSipe,Rubin,ThePhoton}. The photodetection
amplitude has been identified with the real space photon wave function in the
discussion of down conversion experiments \cite{WaveFunction}, a choice that
can be justified by it's relationship to photon counting which can localize
the photon \cite{ScullyBook,ScullyOPN}. In the standard formulation of quantum
mechanics the real space wave function is the projection of the state vector
onto an orthonormal basis of eigenvectors of a Hermitian position operator.
However, it has been claimed since the early days of quantum mechanics that
there is no position operator that defines such a basis for the photon. Here
we will briefly review our recent work on the construction of a photon
position operator and obtain a photon wave function by projecting onto its eigenvectors.

Attempts to arrive at a photon position operator and its associated basis of
localized states go back to the early days of quantum mechanics. Pauli stated
that the nonexistence of a density for the photon corresponds to the fact that
the position of a photon cannot be associated with any operator in the usual
sense \cite{Pauli}. \ Based on definitions of center of mass, Pryce found the
$\mathbf{k}$-space photon position operator $\widehat{\mathbf{r}}_{P}%
=i\nabla-i\mathbf{k/}2k^{2}+\mathbf{k\times S}/k^{2}$ where $S_{j}$ are the
$3\times3$\ spin $1$ matrices, $\mathbf{k}$ is a wave vector, and $\nabla
_{j}=\partial/\partial k_{j}$ \cite{Pryce}. This operator does not have
commuting components, and thus three spatial coordinates cannot simultaneously
have a definite value. In 1949 Newton and Wigner sought \emph{rotationally
invariant} localized states and the corresponding position operators. They
were successful in the case of massive particles and zero mass particles with
spin $0$ and $1/2$, but found for photons ''no localized states in the above
sense exist'' \cite{NewtonWigner}. This result is widely quoted as a proof of
the nonexistence of a photon position operator. It has been proved that there
is no photon position operator with commuting components that transforms as a
vector \cite{Jordan80}.

Recently we have constructed a position operator with commuting components
that is not rotationally invariant \cite{HawtonPO}, does not transform as a
vector \cite{HawtonBaylisPO}, and thus is consistent with the previous work.
Description of a localized state requires a sum over all $\mathbf{k}$ and a
localized photon can have definite spin in the $\mathbf{k}$-direction, that is
it can have definite helicity, but it cannot have definite spin along any
fixed axis. It is the total angular momentum (AM) that has a definite value
along some specified direction in space \cite{CT,HawtonBaylisAM}. The position
eigenvectors are not spherically symmetric, instead they have a vortex
structure as is observed for twisted light \cite{TwistedLight}. Compared to
the Newton Wigner position operators for which transformation of a particle's
spin and position are separable, the photon position operator must incorporate
an additional unitary transfomation that reorients this vortex.

Maxwell's equations are analogous to the Dirac equation when written in terms
of the Riemann-Silberstein field vector $\mathbf{F=E\pm}ic\mathbf{B}$
\cite{BB2,Good} where $\mathbf{E}$ and $\mathbf{B}$ are the electric and
magnetic fields. This suggests that the photon is an elementary particle like
any other, and that Maxwell's equations provide a first quantized description
of the photon. The use of the positive frequency Riemann-Silberstein vector as
a photon wave function has been thoroughly studied \cite{BB,Sipe}. If a field
$\mathbf{\Psi}^{(1/2)}$ such as $\mathbf{F}$\ that goes as $k^{1/2}$ is used
as wave function, a metric factor $k^{-1} $ is required in the scalar product.
The real space squared norm then goes as$\ \int d^{3}r\int d^{3}r^{\prime
}\mathbf{\Psi}^{(1/2)\ast}\left(  \mathbf{r}\right)  \cdot\mathbf{\Psi
}^{(1/2)}\left(  \mathbf{r}^{\prime}\right)  /\left|  \mathbf{r}%
-\mathbf{r}^{\prime}\right|  ^{2}$ and thus its integrand cannot be
interpreted as a local number density \cite{BB2}. Since the photon has no
mass, it has been suggested that there is no photon number density, only
energy density \cite{Sipe}. However, the Landau-Peierls (LP) wave function,
$\mathbf{\Psi}^{(0)}$, whose absolute value squared has been interpreted as
photon number density was investigated as early as 1930
\cite{LandauPeierls,AB}. It has the disadvantage that its relationship to
electric current density and the electromagnetic fields is nonlocal in real
space \cite{BB2,PikeSarkar,Cook}. It is possible to define a biorthonormal
basis with a local scalar product that involves the eigenvectors of an
operator and its adjoint \cite{Fonda}. This formalism has recently been
applied to pseudo-Hermitian Hamiltonians that possess real spectra
\cite{Mostafazadeh}. We will show here that such a basis provides an
interesting alternative to explicit inclusion of a metric operator when
working with electromagnetic fields.

In this letter one and two photon wave functions will be obtained by
projection onto a basis of position operator eigenkets. Our work on the photon
position operator will first be reviewed and the properties of biorthonormal
bases will be outlined. The position eigenkets will be obtained in the
Heisenberg picture (HP). We will then derive photon wave functions from
quantum electrodynamics (QED) by projecting the state vector onto simultaneous
eigenkets of the photon position operator and the number operator.\ Finally we
will discuss the relationship of these projections to the photodetection
amplitude and other real space descriptions of the photon state in the recent literature.

We start with a discussion of the photon position operator. A $\mathbf{k}%
$-space position operator with commuting components and transverse
eigenvectors in the spherical polar $\widehat{\mathbf{\theta}}$ and
$\widehat{\mathbf{\phi}}$ directions was introduced in Ref. \cite{HawtonPO}.
It was generalized in Ref. \cite{HawtonBaylisPO} to allow for rotation about
$\mathbf{k}$ through the Euler angle $\chi\left(  \theta,\phi\right)  $ to
give $\widehat{\mathbf{r}}^{(\alpha,\chi)}=D\left(  k^{\alpha}i\nabla
k^{-\alpha}\right)  D^{-1}$ where $D=\exp\left(  -i\mathbf{S\cdot}%
\widehat{\mathbf{k}}\chi\right)  \exp\left(  -iS_{3}\phi\right)  \exp\left(
-iS_{2}\theta\right)  .$ The unitary transformation $D$ rotates $\mathbf{k}$
from the $z$-axis to an orientation described by the angles $\theta$ and
$\phi,$ while the transverse vectors $\widehat{\mathbf{x}}$ and $\widehat
{\mathbf{y}}$ are rotated first to $\widehat{\mathbf{\theta}}$ and
$\widehat{\mathbf{\phi}}$ and then about $\mathbf{k}$ through $\chi$ to give
the unit vectors
\[
\mathbf{e}_{\mathbf{k},\lambda}^{(\chi)}=\exp\left(  -i\lambda\chi\right)
\left(  \widehat{\mathbf{\theta}}+i\lambda\widehat{\mathbf{\phi}}\right)
/\sqrt{2}%
\]
with helicity $\lambda=\pm1.$ The similarity transformation $k^{\alpha}$
results in eigenkets proportional to $k^{\alpha}$ where we are interested in
$\alpha=0$ and $\pm1/2$ as discussed above. The $\mathbf{k}$-space position
operator,
\begin{equation}
\widehat{\mathbf{r}}^{(\alpha,\chi)}=i\nabla-i\alpha\frac{\mathbf{k}}{k^{2}%
}+\frac{\mathbf{k}\times\mathbf{S}}{k^{2}}\mathbf{-}\frac{\mathbf{\mathbf{k}%
\cdot S}}{k^{2}}\left(  \widehat{\mathbf{\phi}}\cot\theta-\nabla\chi\right)
,\label{r}%
\end{equation}
has transverse $3$-vector eigenkets satisfying
\begin{equation}
\widehat{\mathbf{r}}^{(\alpha,\chi)}\psi_{\mathbf{r}_{1},\lambda_{1}%
,j}^{(\alpha)}\left(  \mathbf{k}\right)  =\mathbf{r}_{1}\psi_{\mathbf{r}%
_{1},\lambda_{1},j}^{(\alpha)}\left(  \mathbf{k}\right) \label{Evect}%
\end{equation}
for a photon with helictiy $\lambda_{1}$ at $\mathbf{r}_{1}.$ In Eq.
(\ref{Evect}) the functional dependence distinguishes $\mathbf{k}$-space from
$\mathbf{r}$-space, while subscripts denote eigenvalues and Cartesian
components. There is a remarkable analogy between the last term in this
$\mathbf{k}$-space position operator and the $\mathbf{r}$-space vector
potential of a magnetic monpole where the Euler angle $\chi$ corresponds to a
change of gauge. This was explored in Ref. \cite{HawtonBaylisPO}. It turns out
that the analogy is primarily mathematical, and the last term in Eq.(\ref{r})
does change the physics. The spin and orbital AM of a photon are not separable
\cite{CT}. However, the $z$-component of the total AM operator commutes with
the position operator and and this allows $\widehat{\mathbf{r}}^{(\alpha
,\chi)},$ the helicity operator $\widehat{\mathbf{k}}\cdot\mathbf{S}$,
$J_{z}=\hbar\left(  -i\mathbf{p}\times\nabla+\mathbf{S}\right)  $ to have
simultaneous eigenvectors with eigenvalues $\mathbf{r}_{1}$, $\lambda_{1},$
and $\hbar j_{z}$ for integral $j_{z}.$ The the first three terms of
$\widehat{\mathbf{r}}^{(1/2,\chi)}$ are the Pryce position operator,
$\widehat{\mathbf{r}}_{P}$, whose components do not commute. The last term
gives the position operator commuting components, dictates that $\widehat
{\mathbf{r}}^{(\alpha,\chi)}$ transform as a vector only for rotations about
the $z$-axis as can be seen from Eq.(67) of Ref. \cite{HawtonBaylisPO}, and
fixes $j_{z}$ for a given $\widehat{\mathbf{r}}^{(\alpha,\chi)}.$ The quantum
numbers $\left\{  \mathbf{r}_{1},\lambda_{1}\right\}  $ index the basis states
for a given $j_{z}.$

A biorthonormal basis of one photon position eigenkets will now be obtained.
For $\alpha=0$ the operator $\widehat{\mathbf{r}}^{(0,\chi)}$ is self adjoint,
has real eigenvalues, and defines a single orthonormal basis as is usual in
quantum mechanics. For fields, $\alpha=1/2$ and the position operator is not
self-adjoint, rather it is pseudo-Hermitian. The biorthonormal pairs,
$\left\{  \psi_{n},\phi_{n}\right\}  $, of eigenkets of a pseudo-Hermitian
operator and its adjoint satisfy \cite{Fonda,Mostafazadeh}
\begin{align}
\widehat{O}\left|  \psi_{n}\right\rangle  & =O_{n}\left|  \psi_{n}%
\right\rangle ,\;\widehat{O}^{\dagger}\left|  \phi_{n}\right\rangle
=O_{n}^{\ast}\left|  \phi_{n}\right\rangle ,\label{Biorthonormal}\\
\widehat{O}^{\dagger}  & =\eta\widehat{O}\eta^{-1},\;\left\langle \phi
_{n}|\psi_{m}\right\rangle =\delta_{n,m},\nonumber\\
\;\sum_{n}\left|  \psi_{n}\right\rangle \left\langle \phi_{n}\right|   &
=\;\sum_{n}\left|  \phi_{n}\right\rangle \left\langle \psi_{n}\right|
=1,\nonumber
\end{align}
where $\eta$ is a metric operator. If $\rho=\sqrt{\eta}$ \ is the positive
square root of $\eta,$ then $\widehat{o}=\rho\widehat{O}\rho^{-1} $ is
Hermitian. To apply this formalism to the photon we take $\eta=k$ and
$\alpha=-1/2.$ Then $\widehat{o}=\widehat{\mathbf{r}}^{(0,\chi)}$ is Hermitian
and the eigenvectors of $\ \widehat{O}=\widehat{\mathbf{r}}^{(-1/2,\chi)}$ and
$\widehat{O}^{\dagger}=\widehat{\mathbf{r}}^{(1/2,\chi)}$ form a biorthogonal
pair that \ go as $1/\sqrt{k}$ and $\sqrt{k}$ as required by QED for the
vector potential and the electromagnetic fields respectively. Eqs.
(\ref{Biorthonormal}) then give Eq. (\ref{Evect}) and
\begin{align}
\widehat{\mathbf{r}}^{(-1/2,\chi)\dagger}  & =k\widehat{\mathbf{r}%
}^{(-1/2,\chi)}k^{-1}=\widehat{\mathbf{r}}^{(1/2,\chi)},\nonumber\\
\sum_{j}\left\langle \psi_{\mathbf{r}_{2},\lambda_{2},j}^{(-\alpha)}%
|\psi_{\mathbf{r}_{1},\lambda_{1},j}^{(\alpha)}\right\rangle  & =\delta
^{3}\left(  \mathbf{r}_{1}-\mathbf{r}_{2}\right)  \delta_{\lambda_{1}%
,\lambda_{2}},\nonumber\\
\sum_{\lambda,j}\int d^{3}r\left|  \psi_{\mathbf{r},\lambda,j}^{(\alpha
)}\right\rangle \left\langle \psi_{\mathbf{r},\lambda,j}^{(-\alpha)}\right|
& =1\label{PhotonBasis}%
\end{align}
where $\delta^{3}$ is the $3$-dimensional Dirac $\delta$-function and we can
interchange $\alpha$ with $-\alpha$. By multiplying Eq. (\ref{Evect}) for
$\widehat{\mathbf{r}}^{(0,\chi)}$ by $\rho^{\mp1}$ where $\rho=k^{1/2}$ and
inserting $\rho^{\pm1}\rho^{\mp1}=1$ between $\widehat{\mathbf{r}}$ and
$\mathbf{\psi}$ to obtain the $\widehat{\mathbf{r}}^{(\pm1/2,\chi)}$
eigenvector equations it can be proved that the real eigenvectors,
$\mathbf{r}_{1},$ are preserved by the similarity transfomation to the
biorthogonal basis.

The time dependence is determined by the Hamiltonian $\widehat{H}+\widehat
{H}_{0}$ with $\widehat{H}=\sum_{\mathbf{k},\lambda}\hbar kca_{\mathbf{k}%
,\lambda}^{\dagger}a_{\mathbf{k},\lambda}$ where the zero point terms
$\widehat{H}_{0}=\sum_{\mathbf{k},\lambda}\hbar kc/2$ which are unaffected by
the photon state will be omitted here. The operator $a_{\mathbf{k},\lambda}$
annihilates a photon with wave vector $\mathbf{k}$ and helicity $\lambda$. The
operators and their eigenkets are time dependent in the HP \cite{Sakurai}.
Using the unitary time evolution operator $U\left(  t\right)  =\exp\left(
-i\widehat{H}t\right)  ,$ the position operator, given by Eq. (\ref{r}) in the
Schr\"{o}dinger picture, becomes $\widehat{\mathbf{r}}_{HP}^{(\alpha,\chi
)}=U^{\dagger}\left(  t\right)  \widehat{\mathbf{r}}^{(\alpha,\chi)}U\left(
t\right)  $ in the HP with eigenkets $U^{\dagger}\left(  t\right)  \left|
\mathbf{r}_{1},\lambda_{1}\right\rangle $ given by
\begin{equation}
\psi_{\mathbf{r}_{1},\lambda_{1},j}^{(\alpha)}\left(  \mathbf{k},t\right)
=k^{\alpha}e_{\mathbf{k},\lambda_{1},j}\exp\left(  -i\mathbf{k\cdot r}%
_{1}+ikct\right)  /\sqrt{V}\label{Evec}%
\end{equation}
in the $\mathbf{k}$-space representation. Equivalently we can describe the $1
$-photon position eigenkets by defining the operators
\begin{equation}
\widehat{\psi}_{\mathbf{r}_{1},\lambda_{1},j}^{(\alpha)}\left(  t\right)
\equiv\sum_{\mathbf{k}}k^{\alpha}e_{\mathbf{k},\lambda_{1},j}a_{\mathbf{k}%
,\lambda_{1}}^{\dagger}\exp\left(  -i\mathbf{k\cdot r}_{1}+ikct\right)
/\sqrt{V}\label{Operator}%
\end{equation}
and the kets
\begin{equation}
\left|  \psi_{\mathbf{r}_{1},\lambda_{1},j}^{(\alpha)}\left(  t\right)
\right\rangle =\widehat{\psi}_{\mathbf{r}_{1},\lambda_{1},j}^{(\alpha)}\left(
t\right)  \left|  0\right\rangle \label{EvecKet}%
\end{equation}
where $\left|  0\right\rangle $ is the vacuum state. The field operators are
$\widehat{\mathbf{E}}=-\partial\widehat{\mathbf{A}}/\partial t\;$and
$\widehat{\mathbf{B}}=\nabla\times\widehat{\mathbf{A}}$ \ where the vector
potential operator in the Coulomb gauge can be written as $\widehat
{\mathbf{A}}_{\mathbf{r}}\left(  t\right)  =\left[  \widehat{\mathbf{A}%
}_{\mathbf{r}}^{(+)}\left(  t\right)  +\widehat{\mathbf{A}}_{\mathbf{r}}%
^{(-)}\left(  t\right)  \right]  /\sqrt{2}\ $with
\[
\widehat{\mathbf{A}}_{\mathbf{r}}^{(+)}\left(  t\right)  =\mathcal{C}%
\sum_{\mathbf{k},\lambda}k^{-1/2}\mathbf{e}_{\mathbf{k}\lambda}a_{\mathbf{k}%
\lambda}\exp\left(  i\mathbf{k\cdot r}-ikct\right)  /\sqrt{V}%
\]
where $\widehat{\mathbf{A}}_{\mathbf{r},\lambda}^{(-)}\left(  t\right)
=\widehat{\mathbf{A}}_{\mathbf{r},\lambda}^{(+)\dagger}\left(  t\right)  ,$
$\mathcal{C}=\sqrt{\hbar/c\epsilon_{0}},$ and $\epsilon_{0}$ the permittivity
and $c$ the speed of light in vacuum$.$ The $1$-photon operators given by
Eq.(\ref{Operator}) are simply related to the vector potential and electric
field operators through $\widehat{\psi}_{\mathbf{r}_{1},\lambda_{1}%
,j}^{(-1/2)}\left(  \mathbf{k},t\right)  =\widehat{A}_{\mathbf{r}_{1}%
,\lambda_{1},j}^{(-)}/\mathcal{C}$ and $\widehat{\psi}_{\mathbf{r}_{1}%
,\lambda_{1},j}^{(1/2)}\left(  \mathbf{k},t\right)  =\widehat{E}%
_{\mathbf{r}_{1},\lambda_{1},j}^{(-)}/\left(  ic\mathcal{C}\right)  .$

A general state vector in which the number of photons and their wave vectors
are uncertain can be expanded as
\begin{align}
\left|  \Psi\right\rangle  & =c_{0}\left|  0\right\rangle +\sum_{\mathbf{k}%
,\lambda}c_{\mathbf{k},\lambda}a_{\mathbf{k},\lambda}^{\dagger}\left|
0\right\rangle \label{StateVector}\\
& +\frac{1}{2!}\sum_{\mathbf{k},\lambda;\mathbf{k}^{\prime},\lambda^{\prime}%
}\sqrt{\mathcal{N}_{\mathbf{k},\lambda;\mathbf{k}^{\prime},\lambda^{\prime}}%
}c_{\mathbf{k},\lambda;\mathbf{k}^{\prime},\lambda^{\prime}}a_{\mathbf{k}%
,\lambda}^{\dagger}a_{\mathbf{k}^{\prime},\lambda^{\prime}}^{\dagger}\left|
0\right\rangle +..\nonumber
\end{align}
where $c_{0}=\left\langle 0|\Psi\right\rangle ,\;c_{\mathbf{k},\lambda}%
\equiv\left\langle 0\left|  a_{\mathbf{k},\lambda}\right|  \Psi\right\rangle
,$ $c_{\mathbf{k},\lambda;\mathbf{k}^{\prime},\lambda^{\prime}}\equiv
c_{\mathbf{k}^{\prime},\lambda^{\prime};\mathbf{k},\lambda}=\left\langle
0\left|  a_{\mathbf{k},\lambda}a_{\mathbf{k}^{\prime},\lambda^{\prime}%
}\right|  \Psi\right\rangle ,$ and $\mathcal{N}_{\mathbf{k},\lambda
;\mathbf{k}^{\prime},\lambda^{\prime}}=1+\delta_{\mathbf{k},\mathbf{k}%
^{\prime}}\delta_{\lambda,\lambda^{\prime}}$. Division by $2!$ corrects for
identical states obtained when the $\left\{  \mathbf{k},\lambda\right\}  $
subscripts are permuted while $\sqrt{\mathcal{N}}/2$ normalizes doubly
occupied states. The \emph{one photon real space wave function}, equal to the
projection of this state vector onto an eigenket of $\widehat{\mathbf{r}}%
_{HP}^{(\alpha,\chi)}$ is
\begin{align}
\Psi_{j}^{(\alpha)}\left(  \mathbf{r},t\right)   & =\sum_{\lambda}\left\langle
\psi_{\mathbf{r},\lambda,j}^{(\alpha)}|\Psi\right\rangle =\sum_{\mathbf{k}%
,\lambda}c_{\mathbf{k},\lambda}\psi_{\mathbf{r},\lambda,j}^{(\alpha)\ast
}\left(  \mathbf{k},t\right) \label{1photonPhi}\\
& =\sum_{\mathbf{k},\lambda}c_{\mathbf{k},\lambda}e_{\mathbf{k,}\lambda
,j}^{\ast}k^{\alpha}\exp\left(  i\mathbf{k\cdot r}-ikct\right)  /\sqrt
{V}\nonumber
\end{align}
where we have used Eqs. (\ref{EvecKet}), (\ref{Operator}) and
(\ref{StateVector}). If $\alpha=0$ this is the LP wave function,
$\mathbf{\Psi}^{(0)}\left(  \mathbf{r},t\right)  $. The vector potential%
\[
\mathbf{A}^{(+)}\left(  \mathbf{r},t\right)  =\mathcal{C}\mathbf{\Psi
}^{(-1/2)}\left(  \mathbf{r},t\right)
\]
determines the positive frequency fields
\begin{align*}
\mathbf{E}^{(+)}\left(  \mathbf{r},t\right)   & =-\frac{\partial}{\partial
t}\mathbf{A}^{(+)}\left(  \mathbf{r},t\right)  =ic\mathcal{C}\mathbf{\Psi
}^{(1/2)}\left(  \mathbf{r},t\right)  ,\\
\mathbf{B}^{(+)}\left(  \mathbf{r},t\right)   & =\nabla\times\mathbf{A}%
^{(+)}\left(  \mathbf{r},t\right)  ,
\end{align*}
which satisfy Maxwell's equations. The photodetection wave function is
$\mathbf{E}^{(+)}\left(  \mathbf{r},t\right)  =\left\langle 0\left|
\widehat{\mathbf{E}}^{(+)}\left(  \mathbf{r},t\right)  \right|  \Psi
\right\rangle $ \cite{ScullyBook}. The scalar product
\begin{align*}
\left\langle \Psi|\Psi\right\rangle  & =\sum_{j}\int d^{3}r\Psi_{j}%
^{(-\alpha)\ast}\left(  \mathbf{r},t\right)  \Psi_{j}^{(\alpha)}\left(
\mathbf{r},t\right) \\
& =\sum_{\mathbf{k},\lambda}\left|  c_{\mathbf{k},\lambda}\right|  ^{2}%
\equiv\left|  c_{1}\right|  ^{2}%
\end{align*}
has a local integrand and $\left|  c_{1}\right|  ^{2}$ is the probability for
$1$-photon in state $\left|  \Psi\right\rangle $.

In $\mathbf{k}$-space the $1$-photon LP and field wave functions identically
predict probability $\left|  c_{\mathbf{k},\lambda}\right|  ^{2}$ to measure
momentum $\hbar\mathbf{k}$. In real space the LP wave function leads to a
positive definite density $\left|  \mathbf{\Psi}^{(0)}\left(  \mathbf{r}%
,t\right)  \right|  ^{2}.$ When using fields $\mathbf{\Psi}^{(0)}$ is replaced
with the biorthonormal pair $\left\{  \mathbf{\Psi}^{(1/2)},\mathbf{\Psi
}^{(-1/2)}\right\}  $. States with definite photon energy or angular momentum
can have a definite $k$ \cite{CT}, and the relationship between their
description in the LP and the biorthogonal bases is trivial, even in real
space. However, for position eigenkets and real space wave functions in
general these two bases are not so simply related. According to the
competeness relation in Eq.(\ref{PhotonBasis}) $\alpha=1/2$ and $-1/2$ can be
interchanged and the two options averaged to give the real density
\begin{align*}
n\left(  \mathbf{r},t\right)   & =\operatorname{Re}\left\{  \mathbf{\Psi
}^{(1/2)\ast}\left(  \mathbf{r},t\right)  \cdot\mathbf{\Psi}^{(-1/2)}\left(
\mathbf{r},t\right)  \right\} \\
& =\operatorname{Re}\left\{  i\epsilon_{0}\mathbf{E}^{(-)}\left(
\mathbf{r},t\right)  \cdot\mathbf{A}^{(+)}\left(  \mathbf{r},t\right)
/\hbar\right\}
\end{align*}
which is local but \emph{not positive definite}, and thus it is not a true
probability density. This can be seen from the following example: If $\left|
\Psi\right\rangle $ is a $1$-photon state that includes only wave vectors
$\mathbf{k}_{1}$ and $\mathbf{k}_{2}$ both with helicity $\lambda$ where
$c_{\mathbf{k}_{1},\lambda}=c_{\mathbf{k}_{2},\lambda}=1/\sqrt{2}$ then
\begin{align*}
n\left(  \mathbf{r},t\right)   & =\frac{1}{2V}\left\{  2+\left(  \sqrt
{\frac{k_{1}}{k_{2}}}+\sqrt{\frac{k_{2}}{k_{1}}}\right)  \right. \\
& \left.  \times\cos\left[  \left(  \mathbf{k}_{1}-\mathbf{k}_{2}\right)
\cdot r-\left(  k_{1}-k_{2}\right)  ct\right]  \right\}  .
\end{align*}
The cosine term can exceed the spatially uniform time independent terms due to
the $\sqrt{k}$ factors, leading to negative values. It gives zero if an
integral over all space or all time is performed, explaining why the scalar
product is unaffected by the similarity transformation. In an experiment that
integrates over a long enough time or a large enough spatial volume, use of
$\mathbf{\Psi}^{(0)}$ and the biorthonormal pair $\mathbf{\Psi}^{(\pm1/2)}$ is
equivalent. The density $i\epsilon_{0}\mathbf{E}^{(-)}\cdot\mathbf{A}%
^{(+)}/2\hbar+c.c$ has appeared before. The classical linear and angular field
momenta can be written as $\sum_{j}\int d^{3}ri\epsilon_{0}E_{j}\widehat
{O}A_{j}/\hbar$ \cite{CT}. This form can be applied to optical beam AM
calculations within the paraxial approximation \cite{vanEnkNienhuis}. The
number operator $\widehat{n}=i\epsilon_{0}\sum_{j}\widehat{E}_{j}%
^{(-)}\widehat{A}_{j}^{(+)}/2\hbar+h.c.$ transforms as the zeroth component of
a four-vector and satisfies a continuity equation \cite{HawtonMelde}. The one
photon density $n\left(  \mathbf{r},t\right)  $ equals $\left\langle
\Psi\left|  \widehat{n}\right|  \Psi\right\rangle $ and its integral over all
space is time independent consistent with the pair $\left\{  \mathbf{\Psi
}^{(1/2)},\mathbf{\Psi}^{(-1/2)}\right\}  $ forming a basis as implied by Eqs.
(\ref{PhotonBasis}). Action, which is of this form, has appeared in
calculations of laser linewidth \cite{Arnaud}.

For two photons we can project $\left|  \Psi\right\rangle $ onto the
$2$-photon real space basis $\widehat{\psi}_{\mathbf{r},\lambda,i}^{(\alpha
)}\left(  t\right)  \widehat{\psi}_{\mathbf{r}^{\prime},\lambda^{\prime}%
,j}^{(\alpha)}\left(  t^{\prime}\right)  \left|  0\right\rangle .$ Use of
Eq.(\ref{Operator}) and $\left[  a_{\mathbf{k},\lambda},a_{\mathbf{k}^{\prime
},\lambda^{\prime}}^{\dagger}\right]  =\delta_{\mathbf{k},\mathbf{k}^{\prime}%
}\delta_{\lambda,\lambda^{\prime}}$ then gives the correlation amplitude
\begin{align}
\Psi_{i,j}^{(\alpha)}\left(  \mathbf{r},\mathbf{r}^{\prime},t,t^{\prime
}\right)   & =\frac{1}{2!V}\sum_{\mathbf{k},\lambda;\mathbf{k}^{\prime
},\lambda^{\prime}}\sqrt{\mathcal{N}_{\mathbf{k},\lambda;\mathbf{k}^{\prime
},\lambda^{\prime}}}c_{\mathbf{k},\lambda;\mathbf{k}^{\prime},\lambda^{\prime
}}k^{\alpha}\left(  k^{\prime}\right)  ^{\alpha}\nonumber\\
& \times\left[  e_{\mathbf{k},\lambda,i}^{\ast}e_{\mathbf{k}^{\prime}%
,\lambda^{\prime},j}^{\ast}e^{i\mathbf{k\cdot r}-ikct}e^{i\mathbf{k}^{\prime
}\mathbf{\cdot r}^{\prime}-ik^{\prime}ct^{\prime}}\right. \nonumber\\
& \left.  +e_{\mathbf{k}^{\prime},\lambda^{\prime},i}^{\ast}e_{\mathbf{k}%
,\lambda,j}^{\ast}e^{i\mathbf{k\cdot r}^{\prime}-ikct^{\prime}}e^{i\mathbf{k}%
^{\prime}\mathbf{\cdot r}-ik^{\prime}ct}\right]  .\label{2photonPhi}%
\end{align}
which becomes a \emph{two photon wave function} if we set $t^{\prime}=t$. It
is a\emph{\ symmetric linear combination of products of one photon wave
functions} in agreement with Refs. \cite{LapaireSipe} and \cite{Raymer}. The
one and two photon amplitudes given by Eqs. (\ref{1photonPhi}) and
(\ref{2photonPhi}) are consistent with the use of the detection amplitude as a
wave function
\cite{Raymer,LapaireSipe,ThePhoton,WaveFunction,ScullyBook,ScullyOPN}. The
''two-photon quantum state in coordinate space'' obtained by taking the
Fourier transform of the $\mathbf{k}$-space probability amplitude in Ref.
\cite{Eberly} is an example of recent use of the $\alpha=0$ LP wave function.
In either case, the sum over all $n$-photon terms for all positions and
helicities provides a complete basis and thus ''encodes the maximum total
knowledge describing the system'' as required by Schr\"{o}dinger and discussed
in Ref. \cite{Rubin}. For example, either form can be used to transform from
the photon number basis to the quadrature basis if all nonzero $n$-photon
amplitudes are known.

In summary, we have reviewed our previous work where it is demonstrated that a
photon position operator does indeed exist. Because photon spin and orbital
angular momentum are inseparable, its eigenvectors have a vortex structure
like twisted light. We projected the QED state vectors onto simultaneous
eigenvectors of this position operator and the number operator in two
different ways: If all $k^{\prime}s$ are weighted equally the Landau-Peierls
wave function is obtained. This gives a positive definite probability density,
but a nonlocal relationship to fields and current sources. A biorthonormal
basis consisting of eigenkets proportional to the vector potential and
electric field results in a real local density, $i\epsilon_{0}\mathbf{E}%
^{(-)}\cdot\mathbf{A}^{(+)}/2\hbar+c.c,$ which is not positive definite. Both
of these wave functions have played a role in recent analyses of two photon
entanglement \cite{Raymer,LapaireSipe,Eberly}. The two photon wave function is
a symmetrized product of one photon wave functions in agreement with Refs.
\cite{Raymer} and \cite{LapaireSipe}. When all photon numbers are allowed for,
either basis provides a complete description of the quantum state of the
electromagnetic field, equivalent to the QED state vector.

\textit{Acknowledgement: }The author acknowledges the financial support of the
Natural Science and Engineering Research Council of Canada.

\end{document}